\RequirePackage{fix-cm}
\documentclass[smallextended]{svjour3}       
\smartqed  
\usepackage{graphicx}
\usepackage{color}
\journalname{Int J Theor Phys}
\begin{document}

\title{Small-$x$ behavior of non-singlet spin structure function and Bjorken sum rule with pQCD correction up to NNLO and Higher twist correction
}


\author{N. M. Nath         \and
        J. K. Sarma 
}

\institute{N. M. Nath \at
              Department of Physics, Rajiv Gandhi University,\\ Rono Hills, Doimukh-791112, Arunachal Pradesh, India \\
              \email{nayanmaninath@gmail.com}\\
              J. K. Sarma \at
              Department of Physics, Tezpur University, Tezpur-784028, Assam, India \\
              \email{jks@tezu.ernet.in}\\
}

\date{Received: date / Accepted: date}

\maketitle

\begin{abstract}

A calculation of the non-singlet part of spin dependent structure function, $xg_1^{NS}(x,Q^2)$ and associated sum rule, the Bjorken Sum rule up to next-next-to-leading order(NNLO) is presented. We use a unified approach incorporating Regge theory and the theoretical framework of perturbative Quantum Chromodynamics.  Using a Regge behaved model with $Q^2$ dependent intercept as the initial input, we have solved the Dokshitzer-Gribov-Lipatov-Altarelli-Parisi (DGLAP) evolution equation up to NNLO at small-$x$ for $xg_1^{NS}(x,Q^2)$ and the solutions are utilised to calculate the polarised Bjorken sum rule(BSR). We have also extracted the higher twist contribution to BSR based on a simple parametrisation. These results for both of $xg_1^{NS}(x,Q^2)$ and BSR, along with higher twist corrections are observed to be consistent with the available data taken from SMC, E143, HERMES, COMPASS and JLab experiments. In addition, our results are also compared with that of other theoretical and phenomenological analysis based on different models and a very good agreement is also observed in this regard. Further a very good consistency between our calculated results and theoretical QCD predictions of BSR is also achieved.
\end{abstract}

\section{Introduction}
Proper understanding of the spin structure of nucleon and associated sum rules is expected to offer an important opportunity to investigate Quantum Chromodynamics(QCD) as a theory of strong interaction and hence these observables have been the active frontiers in recent years \cite{nmn11,nmn12,nmn13,nmn14,nmn15,nmn16,nmn17,nmn171}. Many successful experimental programs of polarized deep-inelastic lepton-nucleon scattering in combination with remarkable theoretical efforts have been devoted in order to elucidate the internal spin structure of the nucleon. Polarized deep inelastic lepton scattering experiment have been carried out at SLAC, CERN, DESY and Jefferson Laboratory(JLab)\cite{nmn11}.  With the advent of dedicated experimental facilities, recent experiments were able to determine the spin structure functions as well as different sum rules over a wide range of $x$ and $Q^2$ with ever increasing precision. Simultaneously, tremendous progress is observed in the field of theoretical investigation in determining and understanding the shape of quarks and gluon spin distribution functions with perturbative QCD, non-perturbative QCD, chiral perturbation theory\cite{chiral}, lattice QCD\cite{lattice}, anti-de Sitter/conformal field theory (AdS/CFT)\cite{ads/cft}, etc., along with different reliable theoretical models. In addition, recently available several dedicated phenomenological works\cite{ppdf01,ppdf02,ppdf03,ppdf1,ppdf2,ppdf3,ppdf4,ppdf5,ppdf6,ppdf7,ppdf8,ppdf9,ppdf10} in extracting polarized parton distribution function(PPDF) as well as spin structure functions from different experiments within NLO and NNLO QCD analysis have also significant contributions towards the understanding of spin structure of the nucleon.

\ In Quantum Chromodynamics, the spin structure function $g_1(x, Q^2)$ is described as Mellin convolutions between parton distribution functions ($\Delta q_i,\Delta g$) and the Wilson coefficients $C_i$ \cite{nmn21}

\begin{eqnarray}
 g_1 (x,Q^2)=\frac{1}{2n_f} \sum_{i=1}^n e_i^2  [C_{NS}\otimes \Delta q_{NS} + C_S \otimes \Delta q_S+2n_f C_g \otimes \Delta g ],\label{eq1}
\end{eqnarray}

\noindent which consists of three parts, non-singlet ($g_1^{NS} (x,Q^2 )=\frac{1}{2n_f} \sum_{i=1}^n e_i^2  [C_{NS} \otimes \Delta q_{NS}  ]$), singlet ($g_1^S (x,Q^2 )=\frac{1}{2n_f} \sum_{i=1}^n e_i^2  [C_S \otimes \Delta q_S  ]$) and gluon ($ \Delta G(x,Q^2 )=\frac{1}{2n_f} \sum_{i=1}^n e_i^2  [2n_f C_g \otimes \Delta g ]$). The $Q^2$ distribution of these spin dependent non-singlet, singlet and gluon distribution functions are governed by a set of integro-differential equations, the Dokshitzer-Gribov-Lipatov-Altarelli-Parisi (DGLAP) equations which are given by \cite{nmn22,nmn23,nmn24,nmn25}

\begin{equation}
    Q^2 \frac{\partial xg_1^{NS}(x,Q^2)}{\partial {\ln Q^2}}= \frac{\alpha (Q^2)}{2\pi} P_{qq}^{NS}(x,Q^2)\otimes xg_1^{NS}(x,Q^2)\label{eq2},
\end{equation}

\begin{eqnarray}
 Q^2\frac{\partial\bigg( \begin{array}{c}
g_1^S(x,Q^2) \\
\Delta G(x,Q^2)
\end{array}\bigg)}{\partial {\ln Q^2}}
 =\bigg(
\begin{array}{c}
P_{qq}^S(x,Q^2)\\
P_{gq}^S(x,Q^2)
\end{array}
\begin{array}{c}
2n_f P_{qg}^S(x,Q^2)\\
P_{gg}^S(x,Q^2)
\end{array}\bigg)
\otimes \bigg(\begin{array}{c}
g_1^S(x,Q^2) \\
\Delta G(x,Q^2)
\end{array}\bigg)
\label{eq3}.
\end{eqnarray}

\noindent Here $P_i(x,Q^2)$ are the polarized splitting functions \cite{nmn25,nmn26,nmn27,NNLOSP}. These equations are valid to all orders in the strong coupling constant $\frac{\alpha(Q^2)}{2\pi}$.

\ Although QCD predicts the $Q^2$ dependence of structure functions in accord with the DGLAP equations but they have limitations on absolute prediction of structure functions. DGLAP equations cannot predict the initial values from which the evolution starts, they can only predict the evolution of structure functions with $Q^2$, once an initial distribution is given. Further, due to its complicated mathematical structure, an exact analytic determination of the structure functions is currently out of reach and one needs to apply approximated methods to arrive on predictions from the DGLAP equation. Accordingly several approximate numerical as well as semi-analytical methods for the solution of DGLAP equation have been discussed considerably over the past years \cite{sol25,sol3,sol19,sol5,sol4,sol2}. In literature there are essentially two main classes of approaches in order to have solutions of DGLAP equations: those that solve the equation directly in $x$-space and those that solve it for Mellin transformations of structure functions and invert the transformation back to $x$-space. The approaches based on Mellin transformation method have been achieved much interest because under Mellin transformation the integro-differential DGLAP equation turns into a continuum of independent matrix differential equations, one for each value of moments($N$), which in turn makes the evolution more efficient numerically. However, in this regard as the Mellin transformation of both the splitting functions and the initial input is required, which may not be possible for all functions, especially if higher-order corrections are included in the equations, therefore it is not possible to have exact solution to DGLAP equation in moment space beyond leading order. In contrast to Mellin space, the $x$-space method is more flexible, since the inputs are only required in $x$-space; however it is generally considered to less efficient numerically, because of the need to carry out the convolution in DGLAP equations. Taking into account the advantage of being greater flexibility, despite the difficulty in obtaining high accuracy, the $x$-space methods have been serving as the basis of many widely used programs HOPPET\cite{HOPPET}, QCDNUM\cite{sol25}, CANDIA\cite{sol2} etc., and being incorporated by the CTEQ\cite{CTEQ}, MRST/MSTW(see \cite{MSTW} and references therein) collaborations. In addition, several numerical and semi-analytical methods have been developed\cite{HOX1,HOX2,HOX3,nmn28,nmn29,NNPDF} and achieved significant phenomenological success.

\ Due to the unavailability of exact analytical way of solving the DGLAP equations, in current analysis this set of equations are solved numerically by using an initial input distribution for the structure function at a fixed $Q^2$, in terms of some free parameters, the parameters are so adjusted that the parametrization best fit the existing data. However, the consideration of a specific parametrization with large number of parameters is potentially a source of bias, i.e. systematic error which is very difficult to control. Furthermore, when a parametrization is fitted  to the data, it is very hard to obtain a determination not only of the best fitting parameters, but also of their errors. Therefore, explorations of the possibility of obtaining accurate solutions of DGLAP evolution equations without an initial input or with initial input, consisting of less number of parameters are always interesting. NNPDF method is one of the most interesting methods which does not require to assume a functional form and it is largely bias free\cite{ppdf1,NNPDF}.

 \ In order to perform a fit, one must start with a particular ansatz for the structure functions at some reference $Q_0^2$. In most of the existing fitting analysis, including those in the experimental papers it has been performed by assuming a simple power behavior based on Regge theory. The small-$x$ behaviour of $g_1^{NS}$ in accord with Regge theory is given by (see \cite{reggem1}and references therein)

\begin{eqnarray}
g_1^{NS}= \gamma_{NS}x^{-\alpha}
 \end{eqnarray}

 \noindent and describes the SLAC\cite{reggemslac1,reggemslac2} experimental data with $\gamma_{NS}=0.14$ and $\alpha=0.5$\cite{reggem2,reggem3}. Although this Regge model seems to legitimate as far the early data are concerned, which were mostly taken at moderate $Q^2$ ($ \approx 10 GeV^2 $) and  $x$ values of around $x \geq 0.01$, but the recent measurement of $g_1^{NS}$ which are available within the small-$x$ interval $0.0001< x <0.01 $ can be described with a single Regge type exchange $g_1^{NS}=Ax^{\alpha}$, in which the intercept has a smooth $Q^2$ dependence and $g_1^{NS}$ varies like $x^{\alpha}$ with $-0.5 \leq \alpha \leq 0$. Similar behaviour was predicted with valon model\cite{ppdf10} and a variation from $-0.13$ to $-0.3$ was obtained within the interval of $Q^2$ from $2 GeV^2$ to $10 GeV^2$. On the other hand, Ref.\cite{reggem4} predicts a behaviour of the type, $g_1^{NS} \simeq \bigg(\frac{Q^2}{x^2}\bigg)^{\Delta_{NS}/2}$, with $\Delta_{NS} = 0.42$ in which the asymptotic scaling of $g_1^{NS}$ depends on only one variable $\frac{Q^2}{x^2}$. In addition, there are several studies on the small-$x$ behaviour of non-singlet part of spin structure function(see for example \cite{reggem}).

 \ In this manuscript we have investigated the small-$x$ behaviour of $g_1^{NS}$ structure function based on a simple Regge ansatz of the type $g_1^{NS}=Ax^{-bt}$ with $Q^2$ dependent intercept. The underlying idea behind the assumption of this type of model is as follows: HERA measurements\cite{nmn210,nmn211} suggest that the behavior of $F_2(x,Q^2)$ structure function at low-$x$ is consistent with a dependence $F_2 (x,Q^2 )=Cx^{-\lambda(Q^2 ) }$, where the coefficient $A$ is independent of $Q^2$ and the exponent, defined by $\lambda(Q^2 )= a \ln⁡ \bigg( \frac{Q^2}{\Lambda^2}\bigg)=at$, is observed to rise linearly with $ \ln⁡ Q^2$. Here $\Lambda$ is the QCD cut off parameter and $t=\ln⁡ \bigg(\frac{Q^2}{\Lambda^2}\bigg)$. Thus we see that the rise of the un-polarized structure function ($F_2 (x,Q^2 )$) is much steeper than that predicted by Regge theory and gets steeper and steeper as $Q^2$ increases. Since this observation it has been the challenging issue to resolve whether the Regge intercepts for $F_2(x,Q^2)$ structure function as well as it's non-singlet, singlet and gluon parts, along with the spin structure functions are $Q^2$ dependent or not. Further, before the observation at HERA, there are several predictions on the $Q^2$ dependency of the Regge intercept\cite{bt2,bt3}. These predictions as well as experimental observations at HERA motivated us to consider the possibility that the Regge behaved non-singlet part, $\frac{1}{x}F_2^{NS}(x,Q^2)$ of $ F_2 (x,Q^2 )$ structure function is  also satisfy a functional behaviour, $F_2^{NS}(x,Q^2)=Ax^{-b\ln(\frac{Q^2}{\Lambda^2})}=Ax^{-bt}$ similar to $F_2 (x,Q^2 )$. Again as both the non-singlet structure functions, $\frac{1}{x}F_2^{NS}(x,Q^2)$ and $g_1^{NS} (x,Q^2)$ are Regge behaved\cite{ppdf10,regge1}, therefore their $x$ dependency will be similar within smaller-$x$ region. Further, in QCD the $Q^2$ behaviour of these structure functions are governed by the same DGLAP equation. Therefore the $x$ and $Q^2$ dependency for both the non-singlet structure functions are similar and in accord with $F_2(x,Q^2)$, and hence $F_2^{NS}(x,Q^2)$, here we assume that the $Q^2$ dependency of the Regge behaved structure function $g_1^{NS}(x,t)$ is dominated only by the intercept and it satisfies a relation of the type  $g_1^{NS} (x, t )=Ax^{-bt}$.

\ In order to investigate the small-$x$ behaviour of $g_1^{NS}(x,t)$ structure function, in this paper firstly we have assumed that $g_1^{NS} (x,t)$ satisfies the Regge like behavior $g_1^{NS} (x, t )=Ax^{-bt}$, with the $Q^2$ dependent intercept  within smaller $x$ region and then using the model as the initial input, we have solved the DGLAP equation analytically in leading order (LO), next-to-leading order (NLO) and next-next-to-leading order (NNLO). We have performed a phenomenological analysis of these solutions in comparison with different experimental measurements\cite{nmn41,nmn42,nmn43,nmn44} as well as the predictions due to different models \cite{ppdf10,nmn51,nmn52,nmn53} and achieved at a very good phenomenological success. The phenomenological success achieved in this regard reflects, on one hand the acceptability of the Regge ansatz in describing the small $x$ behavior of the non-singlet part of spin structure function and on the other hand, the usefulness of the Regge ansatz in evolving the spin structure function, $g_1^{NS} (x,Q^2 )$ in accord with DGLAP equation with a considerable precision within smaller $x$ region.

\ We have utilized the solutions of the DGLAP equations in determining the Bjorken Sum Rule (BSR)\cite{nmn31}. BSR relates the difference of proton and neutron structure functions integrated over all possible values of Bjorken variable, $x$ to the nucleon axial charge $g_A$\cite{nmn32}. As BSR holds the flavour non-singlet combination of the spin structure function, it is not marred by the presence of the sea quark and gluon densities about which we have very poor information in particular in the small-$x$ region, therefore QCD analysis by means of BSR is less complex and more accurate.  At infinite four-momentum transfer squared, $Q^2$, the sum rule reads

\begin{eqnarray}
\Gamma_1^{p-n}= \Gamma_1^p -\Gamma_1^n = \int_0^1\frac{dx}{x}xg_1^{NS}(x,Q^2)=\frac{g_A}{6}\label{eq4}.
 \end{eqnarray}

 However, away from $Q^2\rightarrow \infty$,  the polarized Bjorken sum rule is given by a sum of two series in powers of the strong coupling constant $\alpha_s(Q^2)$ (leading twist pQCD correction)and in powers of $\frac{1}{Q^2}$(nonperturbative higher twist corrections):

\begin{eqnarray}
\Gamma_1^{p-n}(Q^2)=\int_0^1\frac{dx}{x}xg_1^{NS}(x,Q^2)=\frac{g_A}{6} \bigg[1-\frac{\alpha_s}{\pi}-3.583(\frac{\alpha_s}{\pi})^2\nonumber\\-20.215(\frac{\alpha_s}{\pi})^3+.........\bigg] + \sum_{i=2}^\infty\frac{\mu_{2i}^{p-n}(Q^2)}{Q^{2i-2}}\label{eq5}
 \end{eqnarray}

\noindent Here the leading twist term (bracket term) consists of pQCD results up to third order of $\alpha_s(Q^2)$\cite{nmn33}. Recently the calculation of the fourth order contribution to the BSR is also available\cite{nmn34,nmn35}. The second term on r.h.s. is known as higher twist term. The higher order pQCD corrections and higher twist power corrections are significant at low-$Q^2$ region(see Ref. \cite{nmn36,nmn37} and references therein). In this paper we have paid attention to  both the parts. As our analysis is based on the solution of DGLAP equation and the splitting functions for $g_1^{NS} (x,Q^2 )$ are not available beyond NNLO, we could not incorporate the available NNNLO corrections to BSR and restricted our study within NNLO perturbative corrections. Required higher twist corrections to describe low-$Q^2$ data with different perturbative orders are extracted based on a simple parametrisation and by means of fitting with the available experimental data.

\ BSR represents the area under the curve $g_1^{NS} (x,Q^2 )$ from from $x=0$ to $x=1$ and in accord with NNLO pQCD, the $Q^2$ dependence of this area is given by,

\begin{eqnarray}
\Gamma_1^{p-n}(Q^2)=\int_0^1\frac{dx}{x}xg_1^{NS}(x,Q^2)=\frac{g_A}{6} \bigg[1-\frac{\alpha_s}{\pi}-3.583(\frac{\alpha_s}{\pi})^2-20.215(\frac{\alpha_s}{\pi})^3\bigg]\label{eq6}.
 \end{eqnarray}

\noindent Here $g_A=1.2695\pm 0.0029$ is the scale-invariant isovector axial-charge measured in neutron beta-decays\cite{nmn32}. The Bjorken integral can be resolved as

 \begin{equation}
\Gamma_1^{p-n}(Q^2)=\int_0^{_{min}}\frac{xg_1^{NS}(x,Q^2)}{x}dx+\int_{x_{min}}^1\frac{xg_1^{NS}(x,Q^2)}{x}dx \label{eq7}
 \end{equation}

\noindent which gives

 \begin{equation}
\Gamma_1^{p-n}(x_{min},Q^2)=\int_{x_{min}}^1\frac{xg_1^{NS}(x,Q^2)}{x}dx=\Gamma_1^{p-n}(Q^2)-\int_0^{x_{min}}\frac{xg_1^{NS}(x,Q^2)}{x}dx\label{eq8}.
 \end{equation}

\noindent The integral on the left hand side of (\ref{eq8}) represents the area under the curve $\frac{xg_1^{NS} (x,Q^2 )}{x}$  from $x=x_{min}$ to $x=1$. For $x = x_{min}\rightarrow 0$, this integral will tend to cover the whole area under the curve from $x=x_{min}=0$ to $x=1$, that is, it will represent the Bjorken integral. Again the second part  on the right side of (\ref{eq8}) represents the part of total area $\int_0^1\frac{xg_1^{NS}(x,Q^2 )}{x}dx$, laying  under the curve $\frac{xg_1^{NS} (x,Q^2 )}{x}$ within  smaller $x$ region i.e., from $x=0$ to any smaller value $x=x_{min}$. Thus we see that in order to investigate BSR, we just require the knowledge of  $xg_1^{NS} (x,Q^2 )$ structure function within smaller $x$ region. This requirement is fulfilled by using the solutions of DGLAP equations, which provide well description of the small-$x$ behaviour of  $xg_1^{NS}(x,Q^2)$ structure function and determined the $Q^2$ behaviour of BSR. When these results for BSR are compared with  different experimental data \cite{nmn41,nmn42,nmn44,BSRJLAB1,BSRJLAB2,BSRJLAB3} and the predictions made in Ref.~\cite{BSR BI,BSR Soffer,KPSST,KS} along with QCD predictions up to NNLO corrections\cite{nmn33}, we have observed a very good consistency among them.

\ The paper is organized as follows: in the next section, Sec.~\ref{sec:level2} we have discussed about obtaining the small $x$ behavior of $g_1^{NS} (x,Q^2 )$ structure function by solving of the DGLAP equation for $xg_1^{NS} (x,Q^2 )$ in leading order(LO), next-to-leading order(NLO) and next-next-to-leading order(NNLO) using the Regge like ansatz, $g_1^{NS} (x,Q^2 )=Ax^{-b \ln⁡(\frac{Q^2}{\Lambda^2})}$ as the initial input. Also in this section, we have performed a phenomenological analysis of our results in comparison with different experimental data as well as several model predictions. These expressions as the solutions of DGLAP equation in LO, NLO and NNLO are then used to determine the Bjorken integral with QCD corrections up to NNLO in Sec.~\ref{sec:level3.1}, along with a phenomenological analysis of the results. In Sec.~\ref{sec:level3.2}, we have extracted the higher twist terms, associated with different perturbative orders based on a simple parametrisation. Finally, in Sec.~\ref{sec:level4}, we have concluded the paper with a brief discussion.

\section{\label{sec:level2}Small $x$ behavior of $g_1^{NS}(x,Q^2)$ by combining a Regge ansatz and the DGLAP equation}

We have considered that the non-singlet part of the spin structure function satisfies the following Regge ansatz:

\begin{equation}
 g_1^{NS}(x,t)=A.x^{(-bt)}\label{eq251}
\end{equation}

\noindent and hence we have

\begin{equation}
 xg_1^{NS}(x,t)=A.x^{(1-bt)}\label{eq252}.
\end{equation}

For simplicity we denote $xg_1^{NS}(x,t)=g^{NS}(x,t)$ and with this notation, $t$ dependence of $xg_1^{NS}(x,t)$ structure function at a particular value of $x=x_0$ is givent by

\begin{equation}
 g^{NS}(x_0,t)=A.x_0^{(1-bt)}\label{eq26}.
\end{equation}

 Dividing (\ref{eq252}) by (\ref{eq26}) we have the following relation

 \begin{equation}
 g^{NS}(x,t)=g^{NS}(x_0,t)\Bigg(\frac{x}{x_0}\Bigg)^{(1-bt)},\label{eq27}
 \end{equation}

\noindent which gives both  $t$ and $x$ dependence of $g^{NS}(x, t)$ structure function in terms of the $t$ dependent function $g^{NS}(x_0,t)$ at fixed $x=x_0$. The $t$ dependent function, $g^{NS} (x_0,t)$ can be obtained from the DGLAP equation.

\ The DGLAP equation for $xg_1^{NS}(x,Q^2)$ structure function can be written as

\begin{equation}
 \frac{\partial  g^{NS}(x,t)}{\partial t}= \frac{\alpha (t)}{2\pi} \int_{x}^{1} \frac{d\omega}{\omega} g^{NS}\bigg(\frac{x}{\omega},t\bigg) P_{qq}^{NS}(\omega)
 \label{eq9},
\end{equation}

\noindent in terms of the variable $t=\ln(\frac{Q^2}{\Lambda^2})$. Here the splitting function, $P_{qq}^{NS}(\omega)$ is defined up to next-next-to-leading order by $P_{qq}^{NS} (\omega)=  \frac{\alpha (t)}{2\pi} P^0 (\omega)+\bigg(\frac{\alpha (t)}{2\pi}\bigg)^2 P^1 (\omega)+\Big(\frac{\alpha (Q^2)}{2\pi}\Big)^{3} P_i^{2}(\omega)$, where, $P^{(0)} (\omega)$, $P^{(1)} (\omega)$ and $P^{(2)} (\omega)$ are the corresponding LO, NLO and NNLO corrections to the splitting functions\cite{nmn25,nmn26,nmn27,NNLOSP}. Again, in LO, NLO and NNLO the coupling constant $\frac{\alpha (t)}{2\pi}$ is given by $\bigg(\frac{\alpha (t)}{2 \pi}\bigg)_{LO}=\frac{2}{(\beta_0 t)}$, $\bigg(\frac{\alpha (t)}{2 \pi}\bigg)_{NLO}=\frac{2}{\beta_0 t} \bigg[1-\frac{\beta_1  \ln ⁡t}{\beta_0^2 t}\bigg]$ and ${\Bigg(\frac{\alpha (t)}{2\pi}\Bigg)}_{NNLO}=\frac{2}{\beta_0t}\Bigg[1-\frac{\beta_1 \ln t}{\beta_0^2 t} + \frac{1} {\beta_0^2t^2} \Bigg[{\Bigg(\frac{\beta_1}{\beta_0}\Bigg)}^2(\ln^2t-\ln t-1)+\frac{\beta_2}{\beta_0}\Bigg]\Bigg]$ respectively, where $\beta_0$, $\beta_1$ and $\beta_2$ are the corresponding one loop, two loop and three loop corrections of QCD $\beta$ function\cite{alpha}. Substituting the respective splitting functions along with the corresponding running coupling constant in (\ref{eq9}) the DGLAP evolution equations in LO, NLO and NNLO become

\begin{equation}
    \frac{\partial  g^{NS}(x,t)}{\partial t}={\Bigg(\frac{\alpha (t)}{2\pi}\Bigg)}_{LO}\Bigg[\frac{2}{3}\{3+4ln(1-x)\} g^{NS}(x,t)+I_{1}(x,t)\Bigg] \label{eq10},
\end{equation}

\begin{eqnarray}
    \frac{\partial  g^{NS}(x,t)}{\partial t}= {\Bigg(\frac{\alpha (t)}{2\pi}\Bigg)}_{NLO}\Bigg[\frac{2}{3}\{3+4ln(1-x)\} g^{NS}(x,t)+I_{1}(x,t)\Bigg]\nonumber\\+{\Bigg(\frac{\alpha (t)}{2\pi}\Bigg)}^2_{NLO} I_{2}(x,t) \label{eq11},
\end{eqnarray}

\noindent and

\begin{eqnarray}
    \frac{\partial g^{NS}(x,t)}{\partial t}= {\Bigg(\frac{\alpha (t)}{2\pi}\Bigg)}_{NNLO}\Bigg[\frac{2}{3}\{3+4ln(1-x)\}g^{NS}(x,t)+I_{1}(x,t)\Bigg]\nonumber\\+{\Bigg(\frac{\alpha (t)}{2\pi}\Bigg)}^2_{NNLO} I_{2}(x,t)+{\Bigg(\frac{\alpha (t)}{2\pi}\Bigg)}^3_{NNLO} I_{3}(x,t) \label{eqnnlo}
\end{eqnarray}

\noindent respectively.  Here the integral functions are given by

\begin{equation}
I_{1}(x,t)= \int_{x}^{1} \frac{d\omega}{1-\omega}\bigg\{\frac{(1+\omega^2)}{\omega}g^{NS}\Bigg(\frac{x}{\omega},t\Bigg)-2g^{NS}(x,t)\Bigg\} \label{eq12},
\end{equation}

\begin{equation}
I_{2}(x,t)= \int_{x}^{1} \frac{d\omega}{{\omega}} P^{(2)}(\omega)g^{NS}\Bigg(\frac{x}{\omega},t\Bigg) \label{eq13}.
\end{equation}

\noindent and

\begin{equation}
I_{3}(x,t)= \int_{x}^{1} \frac{d\omega}{{\omega}} P^{(3)}(\omega)g^{NS}\Bigg(\frac{x}{\omega},t\Bigg) \label{eqNNLO1}.
\end{equation}

\ We now solve these DGLAP evolution equations in LO, NLO and NNLO using the $Q^2$ dependent Regge ansatz, $g^{NS}(x,t) = xg_1^{NS}(x,Q^2 ) = Ax^{1-b \ln⁡(\frac{Q^2}{\Lambda^2})}= Ax^{1-bt}$. Substituting $g^{NS}(x,t) = Ax^{1-b t}$ and $g^{NS}(\frac{x}{\omega},t) = \omega^{-(1-bt)} g^{NS}(x,t)$ in equation (\ref{eq10}) and rearranging a bit we can convert the LO DGLAP equation into an ordinary differential equation which can be easily solved to have

\begin{equation}
g^{NS}(x,t)\bigg |_{LO}=C\exp\Bigg[\int{\Bigg(\frac{\alpha (t)}{2\pi}\Bigg)}_{LO}P(x,t)dt\Bigg]. \label{eq16}
\end{equation}

\noindent Here

\begin{eqnarray}
P(x,t)=\frac{2}{3}\{3+4ln(1-x)\}
+\frac{4}{3}\int_{x}^{1} \frac{d\omega}{1-\omega}\Bigg\{\frac{1+\omega^2}{\omega}\omega^{-(1-b t)}-2\Bigg\}, \label{eq18}
\end{eqnarray}

\noindent and $C$ is the constant of integration.

\noindent At a fixed value of $x=x_0$, the $t$ dependence of the structure function in LO is given by

\begin{equation}
g^{NS}(x_0,t)=C\exp\Bigg[\int{\Bigg(\frac{\alpha (t)}{2\pi}\Bigg)}_{LO}P(x_0,t)dt\Bigg]. \label{eq20}
\end{equation}

\noindent Again the value of the structure function at $x=x_0$ and $t=t_0$ in accord with (\ref{eq16}) is given by

\begin{equation}
g^{NS}(x_0,t_0)=C\exp\Bigg[\int\frac{\alpha (t)}{2\pi}P(x_0,t)dt\Bigg]\Bigg|_{t=t_0}. \label{eq22}
\end{equation}

\noindent Dividing (\ref{eq20}) by (\ref{eq22}) and rearranging a bit we obtain the  $t$ dependence of $g^{NS} (x_0,t)$ in accord with LO DGLAP evolution equation with respect to the point $g^{NS}(x_0,t_0 )$ as

\begin{equation}
g^{NS}(x_0,t)=g^{NS}(x_0,t_0)\exp\Bigg[\int_{t_0}^{t}{\Bigg(\frac{\alpha (t)}{2\pi}\Bigg)}_{LO}P(x_0,t)dt\Bigg]. \label{eq23}
\end{equation}

\ Now substituting $g^{NS}(x_0,t)\bigg|_{LO}$ from (\ref{eq23}) in (\ref{eq27}), we have a relation representing both $x$ and $t$ dependence of structure function in LO, in terms of the input point $g^{NS}(x_0, t_0)$ given by

 \begin{eqnarray}
 g^{NS}(x,t)\bigg|_{LO}=g^{NS}(x_0,t_0)\exp\Bigg[\int_{t_0}^{t}{\Bigg(\frac{\alpha (t)}{2\pi}\Bigg)}_{LO}P(x_0,t)dt\Bigg]\Bigg(\frac{x}{x_0}\Bigg)^{(1-bt)}.\label{eq28}
 \end{eqnarray}

\noindent Proceeding in the similar way we can obtain the relation for $g^{NS}(x, t)$ structure function in NLO and NNLO as

 \begin{eqnarray}
 g^{NS}(x,t)\bigg|_{NLO}=g^{NS}(x_0, t_0)\exp\Bigg[\int_{t_0}^{t}{\Bigg(\frac{\alpha (t)}{2\pi}\Bigg)}_{NLO}P(x_0,t)dt \nonumber\\+\int_{t_0}^{t}{\Bigg(\frac{\alpha (t)}{2\pi}\Bigg)}^2_{NLO}Q(x_0,t)dt\Bigg]\Bigg(\frac{x}{x_0}\Bigg)^{(1-bt)},\label{eq29}
 \end{eqnarray}

 \noindent and

 \begin{eqnarray}
 g^{NS}(x,t)\bigg|_{NNLO}=g^{NS}(x_0, t_0)\exp\Bigg[\int_{t_0}^{t}{\Bigg(\frac{\alpha (t)}{2\pi}\Bigg)}_{NNLO}P(x_0,t)dt\nonumber\\ +\int_{t_0}^{t}{\Bigg(\frac{\alpha (t)}{2\pi}\Bigg)}^2_{NNLO}Q(x_0,t)dt  \nonumber\\ +\int_{t_0}^{t}{\Bigg(\frac{\alpha (t)}{2\pi}\Bigg)}^3_{NNLO}R(x_0,t)dt \Bigg]\Bigg(\frac{x}{x_0}\Bigg)^{(1-bt)}\label{eqNNLOSOL}
 \end{eqnarray}

 \noindent respectively, where

\begin{eqnarray}
 Q(x,t)= \int_{x}^{1} \frac{d\omega}{\omega}P^{(1)}(\omega){\omega}^{-(1-b t)}\label{eq19}
\end{eqnarray}

\noindent and

\begin{eqnarray}
 R(x,t)= \int_{x}^{1} \frac{d\omega}{\omega}P^{(2)}(\omega){\omega}^{-(1-b t)}.\label{eqNNLO19}
\end{eqnarray}

\definecolor{gold}{rgb}{0.9,0.6,0}
\definecolor{lightblue}{rgb}{0,0.7,1}

\begin{figure} 
\centering 


\includegraphics[width=1.0\textwidth]{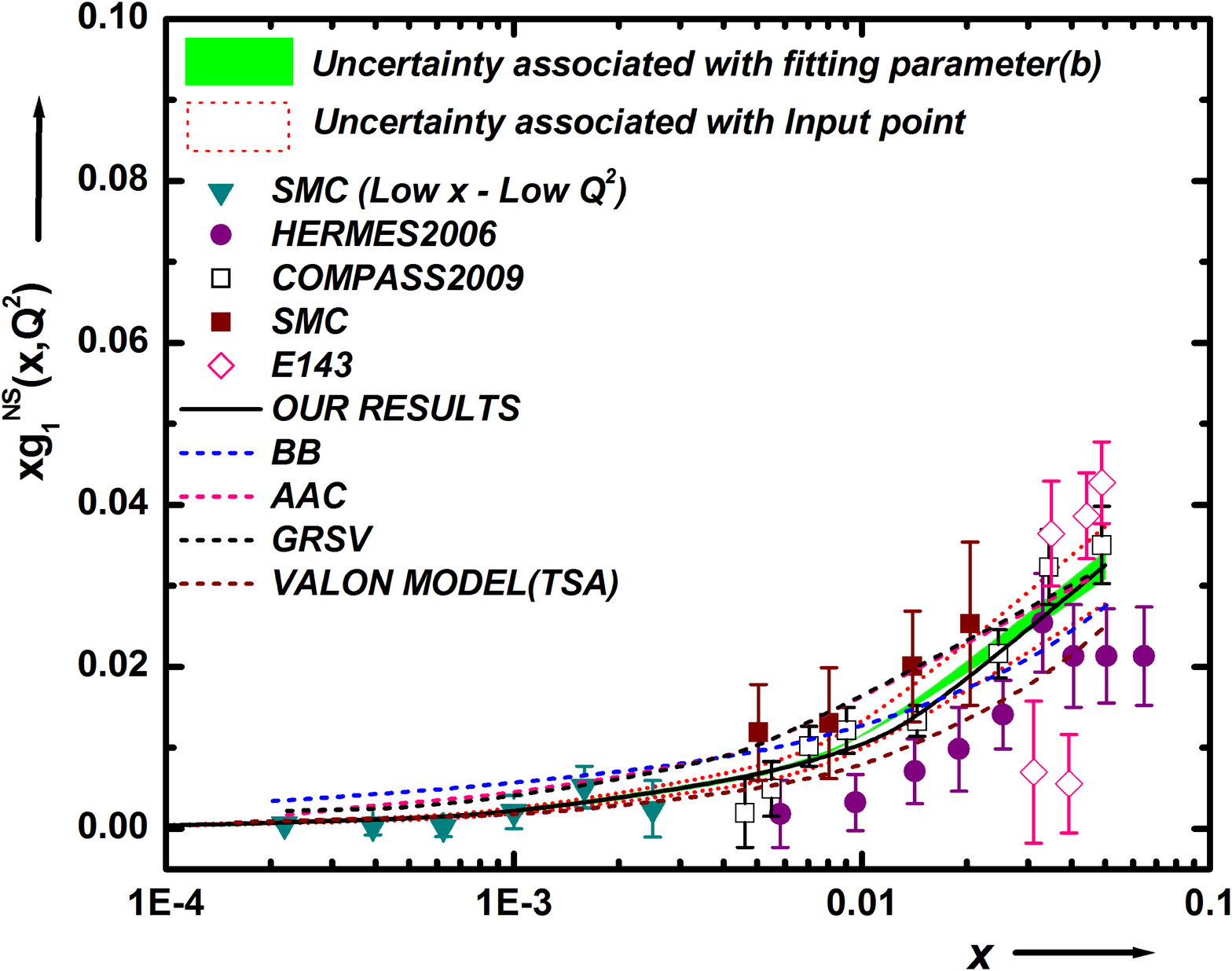}
\caption{$xg_1^{NS}$ structure function at $Q^2 = 5 GeV^2$ compared with the experimental data Ref.~\cite{nmn41,nmn42,nmn43,nmn44} and the results from the analysis presented in Ref.~\cite{ppdf10,nmn51,nmn52,nmn53}.}
\label{fig:fig1}
\end{figure}

\ The expressions, (\ref{eq28}), (\ref{eq29}) and (\ref{eqNNLOSOL}) represent both $x$ and $t$ dependence of $g^{NS}(x,t)$ structure function in LO, NLO and NNLO at small $x$.   They are consisting of the fitting parameter $b$, and a known input point $g^{NS}(x_0, t_0)$ which can be taken from the available experimental data. If the input point is more accurate and precise, we can expect better fitting.  There are not any specific reason in choosing the input point. Any one of the data points at a certain value of $x=x_0$ and $t=t_0$ can be considered as the input point. Of course, the sensitivity of different inputs will be different.  However instead of choosing the input point on the basis of their sensitivity, in our manuscript we have incorporated a suitable condition in determining the input point. We have considered that particular point from the most recent measurements as the input point in which experimental errors are minimum. Under this condition we have selected the point $g_1^{NS} (x_0,t_0 )=0.0133075 \pm 0.001938$ at $x_0=0.0143955$ and $Q^2=5 GeV^2$ from the recent measurement COMPASS. The value of the parameter $b$ is obtained by fitting the parametrization $g^{NS} (x,t)=g^{NS} (x_0,t_0 )  x^{(1-bt)}  x_0^{-(1-bt_0 ) }$, which is a reduced form of $g^{NS} (x,t)=A  x^{(1-bt)}$,  with different experimental data taken from COMPASS\cite{nmn41}, HERMES\cite{nmn42}, SMC\cite{nmn43} and E143\cite{nmn44} collaborations. The parametrization in its reduced form consists of only one fitting parameter $b$ and hence makes easier the fitting analysis. As we are interested in the small $x$ region, therefore we have restricted our fitting analysis with experimental data sets for $x<0.1$ and in this analysis the best fitting value for the parameter $b$ is found to be $b=0.0759\pm 0.0107$ with $\chi^2 /d.o.f.= 1.41$.   Using this value for $b$ we have calculated the $g^{NS}(x, Q^2)$ structure function considering the input point $g^{NS}(x_0=0.0143955,Q^2_0=5 GeV^2)=0.0133075$, and depicted their $x$ evolution for $Q^2=5 GeV^2$ in Fig.~\ref{fig:fig1}, in comparison with SMC, E143, HERMES and COMPASS experimental results, along with the predictions in Ref~\cite{ppdf10,nmn51,nmn52,nmn53} based on various models.Also we have estimated the uncertainty associated with the fitting parameter $b$ and the chosen input point and the respective uncertainty bands are shown in Fig.~\ref{fig:fig1} separately. Here the uncertainty due to the fitting parameter is considerably less than that of due to input point. However both the uncertainties are observed to be decreasing as $x$ decreases. We see that $g^{NS}(x,Q^2)$ structure functions evolved with respect to the input point are consistent with those of experimental measurements as well as other models. This implies that the expressions, we have obtained by means of solving the DGLAP equations analytically, are applicable in describing small $x$ behaviour of $xg_1^{NS}(x,Q^2)$ structure function with a considerable precision and therefore these expressions can be successfully incorporated in $\int_0^{x_{min}}\frac{xg_1^{NS}(x,Q^2)}{x}dx$ for $xg_1^{NS} (x,Q^2 )$ term in order to determine BSR.

\section{\label{sec:level3}Determination of Bjorken Sum Rule}

\subsection{\label{sec:level3.1}Perturbative QCD corrections}

\begin{figure*} 
\centering 
\includegraphics[width=1.0\textwidth]{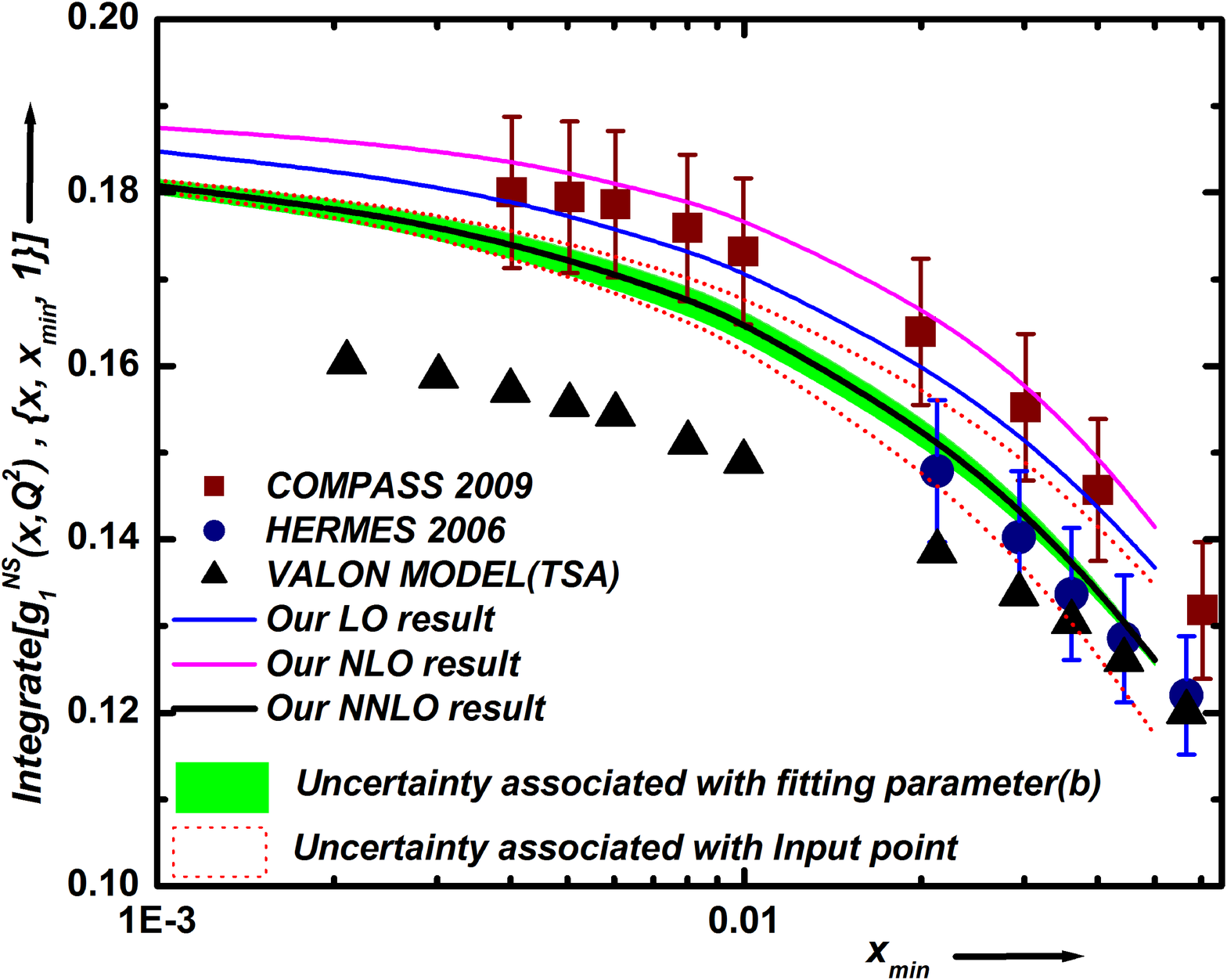}

\caption{ The results of Bjorken integral as a function of the low $x$ limit of integration, $x_{min}$, in LO, NLO and NNLO in comparison with COMPASS \cite{nmn41} and HERMES\cite{nmn42} experimental data  along with the predictions based on Valon model\cite{ppdf10}.}
\label{fig:fig2}
\end{figure*}

\begin{figure*} 
\centering 
\includegraphics[width=1.0\textwidth]{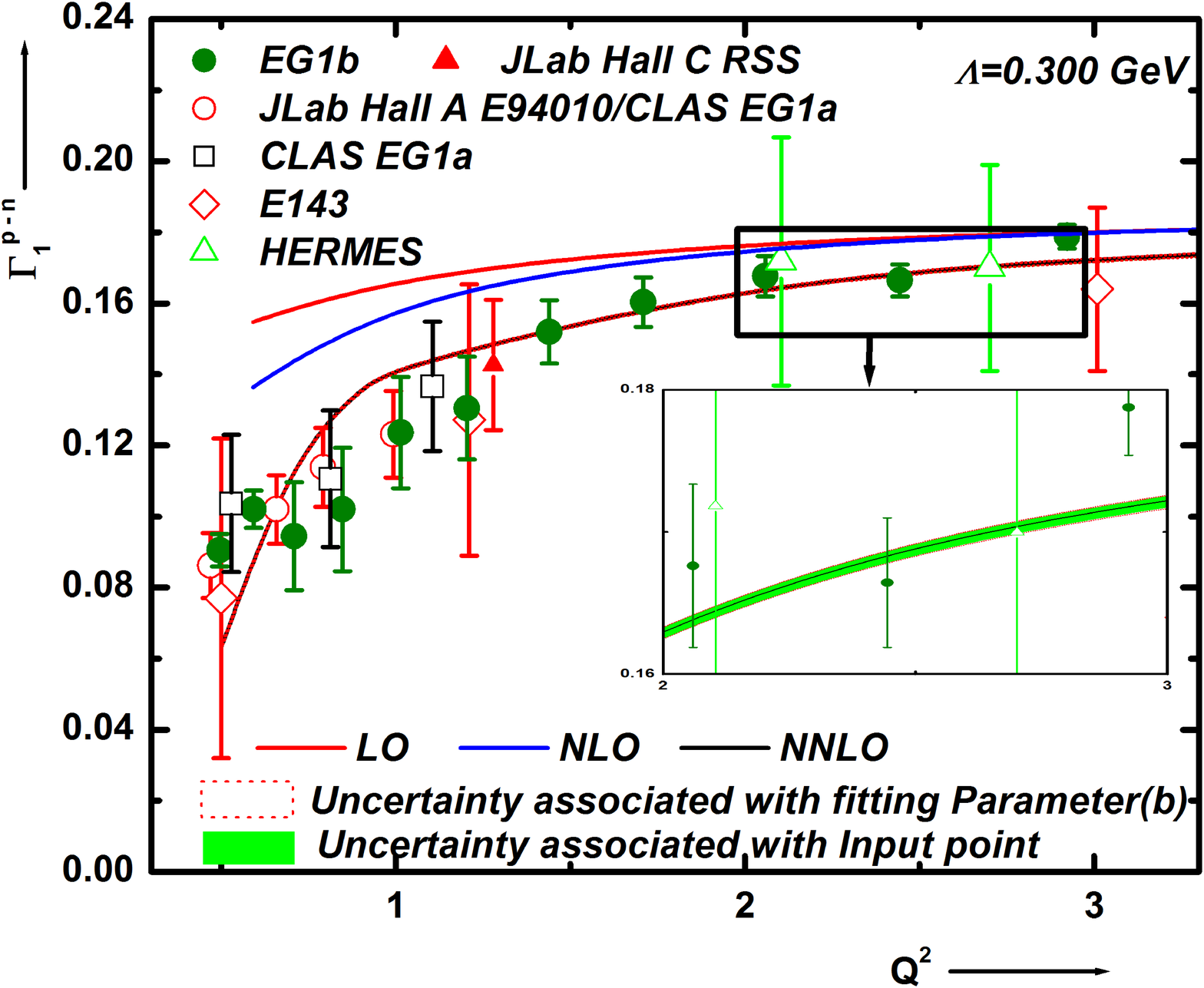}

\caption{ The results of Bjorken integral as a function of momentum transfer squared $Q^2$ in LO, NLO and NNLO against COMPASS \cite{nmn41} and HERMES\cite{nmn42} E143\cite{nmn44} and JLab \cite{BSRJLAB1,BSRJLAB2,BSRJLAB3} experimental data.}
\label{fig:fig3}
\end{figure*}

\begin{figure*} 
\centering 
\includegraphics[width=1.0\textwidth]{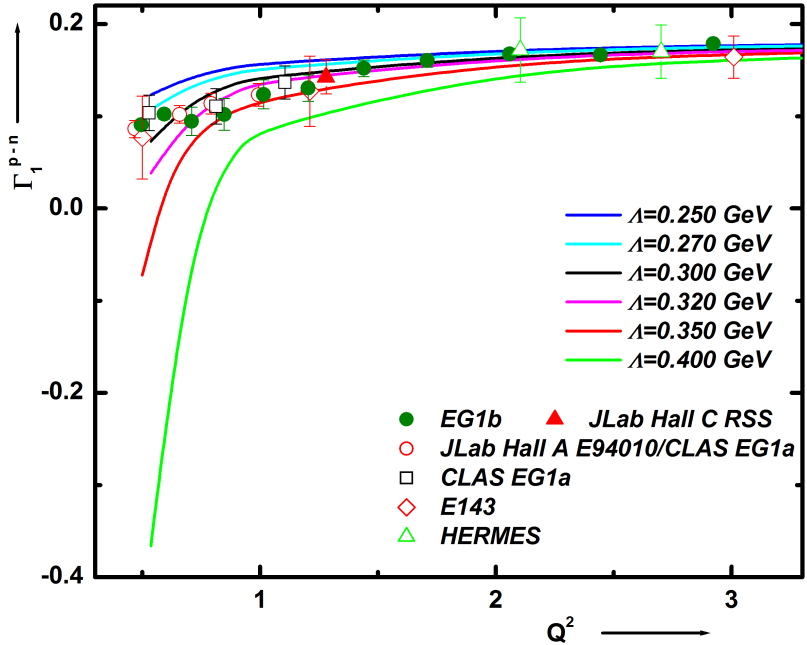}

\caption{$\Lambda$ dependence of our NNLO results for BSR along with COMPASS \cite{nmn41}, HERMES\cite{nmn42}, E143\cite{nmn44} and JLab \cite{BSRJLAB1,BSRJLAB2,BSRJLAB3} experiments.}
\label{fig:fig4}
\end{figure*}

\begin{figure*} 
\centering 
\includegraphics[width=1.0\textwidth]{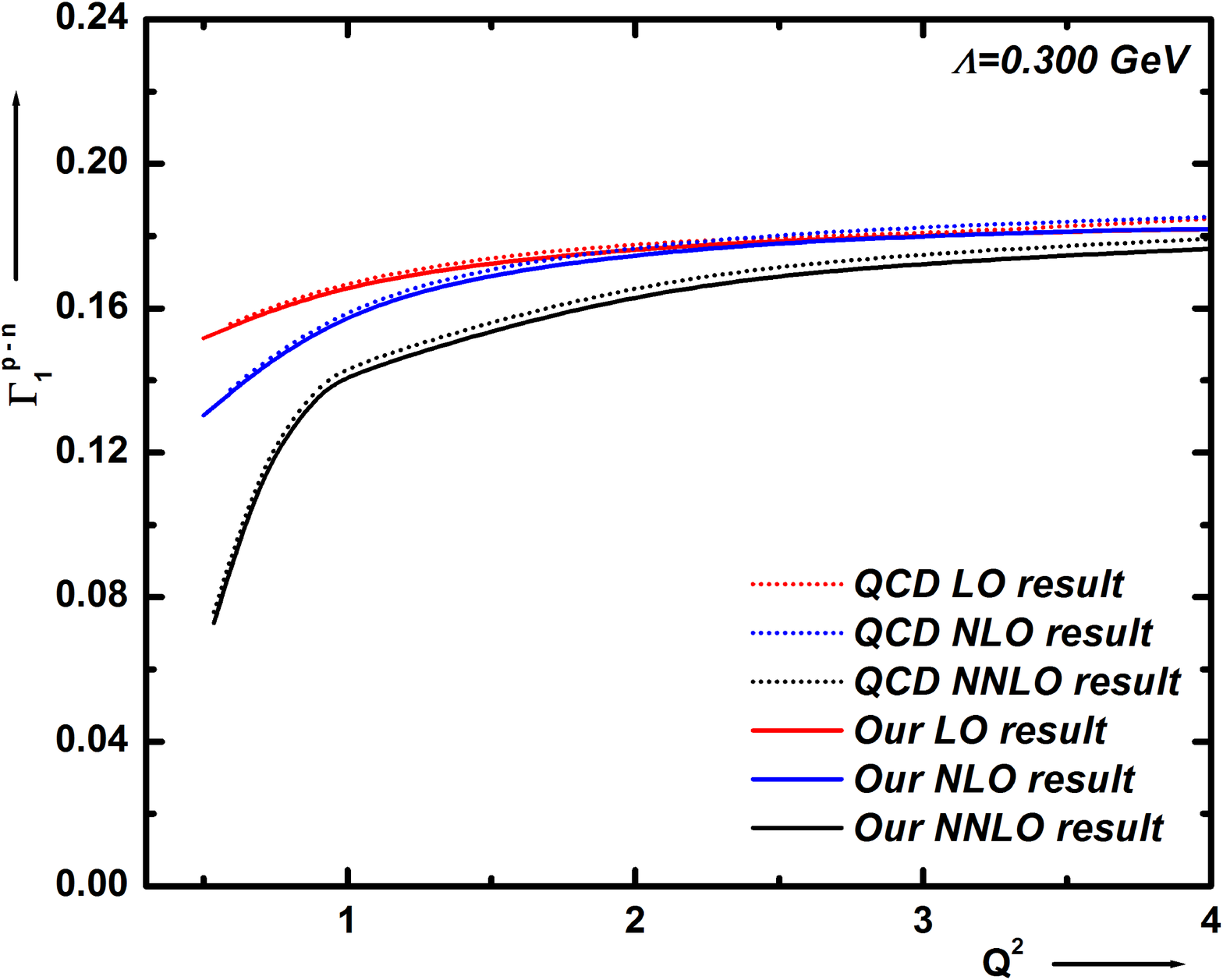}

\caption{Our results of Bjorken integral in comparison with the QCD predictions up to NNLO\cite{nmn33}.}
\label{fig:fig5}
\end{figure*}

The expression (\ref{eq8}) suggest that in order to investigate Bjorken Sum Rule ($\int_{x_{min}\rightarrow 0}^1\frac{xg_1^{NS}(x,Q^2)}{x}dx$), we require the knowledge of the non-singlet spin structure function $xg_1^{NS}(x,Q^2)$ within smaller $x$ region. This requirement can be fulfilled by using the solutions of DGLAP equations obtained above. Therefore, substituting (\ref{eq28}), (\ref{eq29}) and (\ref{eqNNLOSOL})  in (\ref{eq8}) and using the corresponding expressions for $\Gamma_1^{p-n}$ in LO, NLO and NNLO,  we obtain the Bjorken integral with LO, NLO and NNLO QCD corrections as

 \begin{eqnarray}
\Gamma_1^{p-n}(x_{min},Q^2)\bigg|_{LO}=\bigg(BSR(Q^2)\bigg)_{LO}-\int_0^{x_{min}}\frac{dx}{x}\Bigg[g^{NS}(x_0,t_0)\nonumber\\\Bigg(\frac{x}{x_0}\Bigg)^{(1-bt)}\exp\bigg\{\int_{t_0}^{t}{\Bigg(\frac{\alpha (t)}{2\pi}\Bigg)}_{LO}P(x_0,t)dt\Bigg\}\bigg],\label{eq30}
 \end{eqnarray}

 \begin{eqnarray}
\Gamma_1^{p-n}(x_{min},Q^2)\bigg|_{NLO}=\bigg(BSR(Q^2)\bigg)_{NLO}-\int_0^{x_{min}}\frac{dx}{x}\Bigg[g^{NS}(x_0, t_0)\nonumber\\\Bigg(\frac{x}{x_0}\Bigg)^{(1-bt)}\exp\bigg\{\int_{t_0}^{t}{\Bigg(\frac{\alpha (t)}{2\pi}\Bigg)}_{NLO}P(x_0,t)dt\nonumber\\+\int_{t_0}^{t}{\Bigg(\frac{\alpha (t)}{2\pi}\Bigg)}^2_{NLO}Q(x_0,t)dt\bigg\}\Bigg]\label{eq31}
 \end{eqnarray}

\noindent and

 \begin{eqnarray}
\Gamma_1^{p-n}(x_{min},Q^2)\bigg|_{NNLO}=\bigg(BSR(Q^2)\bigg)_{NNLO}-\int_0^{x_{min}}\frac{dx}{x}\Bigg[g^{NS}(x_0, t_0)\nonumber\\\Bigg(\frac{x}{x_0}\Bigg)^{(1-bt)}\exp\bigg\{\int_{t_0}^{t}{\Bigg(\frac{\alpha (t)}{2\pi}\Bigg)}_{NNLO}P(x_0,t)dt\nonumber\\+\int_{t_0}^{t}{\Bigg(\frac{\alpha (t)}{2\pi}\Bigg)}^2_{NNLO}Q(x_0,t)dt+\int_{t_0}^{t}{\Bigg(\frac{\alpha (t)}{2\pi}\Bigg)}^3_{NNLO}R(x_0,t)dt\bigg\}\Bigg]\label{eq31NNLO}
 \end{eqnarray}

\noindent respectively. Considering a known input point $g^{NS}(x_0, t_0)$ from experimental data, we will be able to calculate the BSR integral up to NNLO corrections using the expressions, (\ref{eq30}), (\ref{eq31}) and (\ref{eq31NNLO}) respectively. In our calculations we have used $g^{NS}(x_0=0.0143955,Q^2_0=5 GeV^2)=0.0133075$ as the input point, which is taken from the COMPASS\cite{nmn41} experimental data. With this input point we have calculated the Bjorken integral and the results in accord with equations (\ref{eq30}), (\ref{eq31}) and (\ref{eq31NNLO}) are depicted in Fig.~\ref{fig:fig2} and Fig.~\ref{fig:fig3}.

\ In Fig.~\ref{fig:fig2}, we have plotted our results for BSR integral in LO, NLO and NNLO as a function of low $x$ limit of integration $x_{min}$, in comparison with COMPASS and HERMES measurements along with the results due to valon model(TSA)\cite{ppdf10}. The uncertainties due to the parameter, $b$ and the input point are estimated only for the NNLO results and as seen from the Fig.~\ref{fig:fig2}, they decrease with decrease in $x_{min}$. From Fig.~\ref{fig:fig2} we observe an overall better description of both COMPASS and HERMES data by our results with respect to the predictions due to valon model.  Again our approach expects better results for $x_{min} \rightarrow 0$, but there are no COMPASS measurement for $x < 0.004$ and HERMES measurement beyond $x < 0.02$ for our comparative analysis. Saturation of the COMPASS data for BSR is observed within $x>0.004$, however available HERMES results have not saturated within $x \approx 0.01-0.02$. Thus we may expect to occur saturation within the smaller $x$ region and within this region both HERMES and COMPASS results might agree with each other and  reach an overall compatibility with our measurements.

 \ The $Q^2$ dependency of Bjorken Sum Rule, as predicted by our expressions (\ref{eq30}), (\ref{eq31}), and (\ref{eq31NNLO}) is depicted in Fig.~\ref{fig:fig3}. Here our results are compared with different experimental data taken from COMPASS \cite{nmn41}, HERMES\cite{nmn42}, E143\cite{nmn44} and JLab experiments \cite{BSRJLAB1,BSRJLAB2,BSRJLAB3} and with the theoretical as  well as phenomenological analysis, Ref.~\cite{BSR BI,BSR Soffer,KPSST,KS}. The results depicted in this figure are calculated using the value of $\Lambda = 0.300 GeV$. Here we have also estimated the uncertainty associated with the NNLO results due to the fitting parameter, $b$ and the input point, and they are observed to be very small in this regard. It is also observed that the uncertainty decreases with decrease in $Q^2$. In addition, the dependence of our results, specifically the NNLO results with the QCD parameter $\Lambda$ is shown in Fig.~\ref{fig:fig4}. Fig.~\ref{fig:fig4} reflects that towards low-$Q^2$ region, our results are highly sensitive to $\Lambda$, however the sensitivity decreases towards the smaller values of $\Lambda$. As seen from this figure our approach has the capability for better description of the low-$Q^2$ BSR data with lower values of $\Lambda$.

 \ In Fig.~\ref{fig:fig5}, we  have compared our results with theoretical pQCD predictions (\ref{eq6}) for Bjorken integral up to NNLO. Here our results are calculated with $\Lambda =0.300 GeV$. Within the estimated uncertainty our results show a very good consistency with those of pQCD predictions.

\subsection{\label{sec:level3.2}Higher Twist correction}

We now incorporate the higher twist contribution to our results presented above. Here we have extracted the effect of first non-leading twist term, ($\mu_4$) in LO, NLO and NNLO perturbative orders. To extract the first higher twist term we parameterise the Bjorken sum rule as

\begin{eqnarray}
\Gamma_1^{p-n}(Q^2)= \Gamma_1^{p-n}(Q^2)\bigg|_{LT}+ \frac{\mu_4}{Q^2}\label{eqmu4}
 \end{eqnarray}

\begin{figure} 
\centering %
\includegraphics[width=1.0\textwidth]{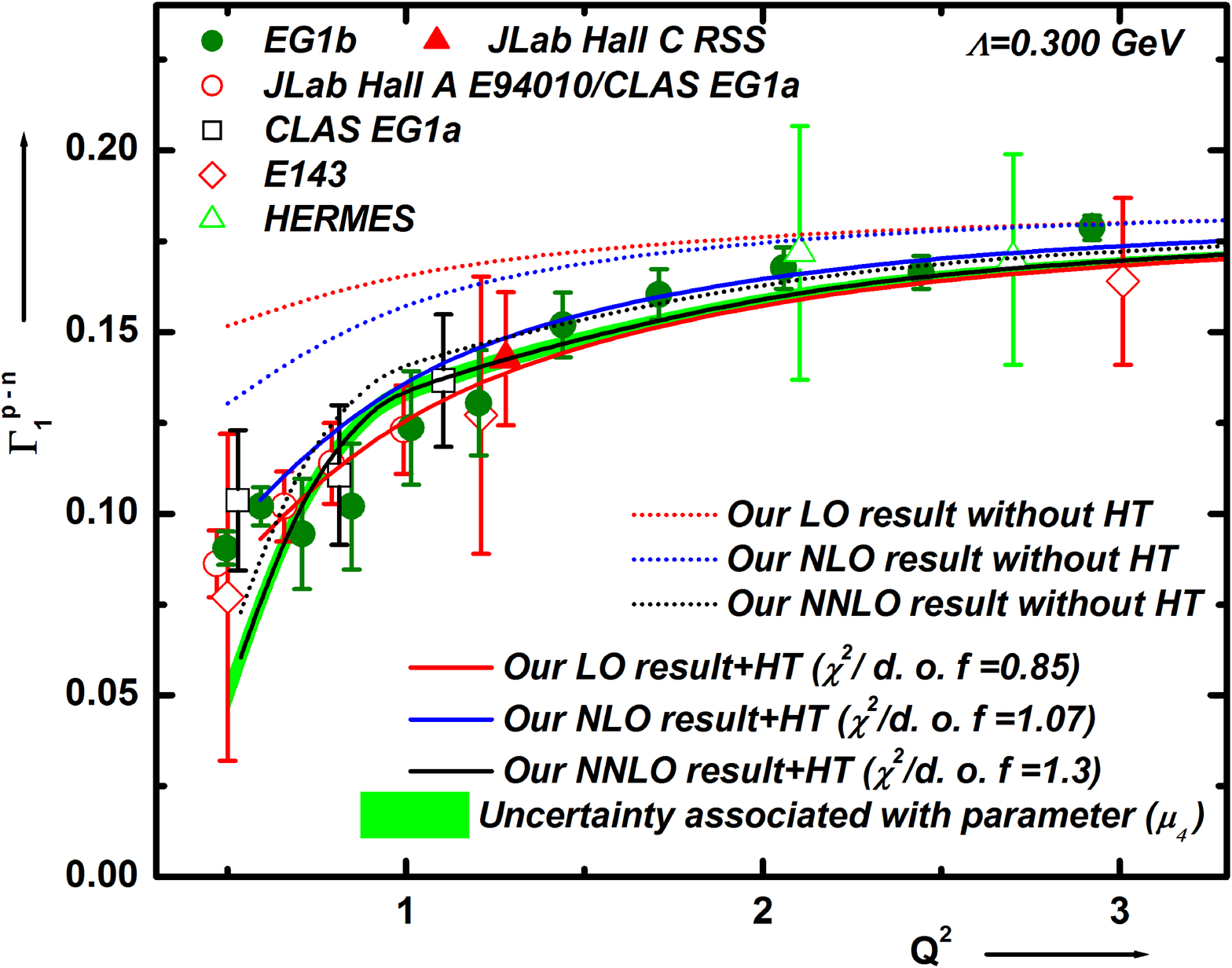}

\caption{Best fitted results for (\ref{eqmu4}) of BSR to the experimental data taken from COMPASS \cite{nmn41}, HERMES\cite{nmn42}, E143\cite{nmn44} and JLab \cite{BSRJLAB1,BSRJLAB2,BSRJLAB3}.}

\label{fig:fig6}
\end{figure}

\noindent Here leading twist(LT) term corresponds to the pQCD contribution to BSR. Incorporating our $Q^2$ dependent expressions  (\ref{eq30}), (\ref{eq31}) and (\ref{eq31NNLO}) for BSR in LO, NLO and NNLO as the LT terms we have fitted above parametrisation to the low $Q^2$ ($0.5\leq Q^2 \leq 5 GeV^2$) experimental data taken from COMPASS \cite{nmn41}, HERMES\cite{nmn42}, E143\cite{nmn44} and JLab experiments \cite{BSRJLAB1,BSRJLAB2,BSRJLAB3}. The corresponding values of $\mu_4$ for which best fitting is obtained in different PT orders are summarised in Table \ref{tab:HT}, along with the respective $\frac{\chi^2} {d.o.f}$ values. The table reflects a substantial change in $\mu_4$-values with different PT orders, which can be interpreted as the manifestation of duality between higher orders and higher twist\cite{DUALITY}.  In Fig.~\ref{fig:fig6} we have presented the best fitting results in various orders of PT. Here both the results, with HT and without HT are shown. We observe that our expressions along with the HT corrections provide well description of BSR data.

\begin{table}[h]
\centering
\begin{tabular}{|l | l | l| l |}
\hline
  & LO & NLO & NNLO \\
\hline
 $\mu_4$ & $-0.034\pm 0.0018$ & $-0.018\pm 0.0014$ & $-0.007\pm 0.0024$ \\
 \hline
 $\frac{\chi^2}{d.o.f}$ & 0.85 & 1.07 & 1.3\\
 \hline

\end{tabular}
\caption{Higher Twist corrections in various Perturbative orders}
\label{tab:HT}
\end{table}

\ The calculations, in this paper are made by using MATHEMATICA 9\cite{mathematica}. In order to obtain sufficient numerical accuracy caution has been taken in evaluation of integrals containing higher order polynomials. The evaluations have been performed using Horner's method and it has improved speed and stability for numeric evaluation of large polynomials present in the splitting functions.

\section{\label{sec:level4}Discussion and conclusions}

In this paper we have obtained some expressions for the non-singlet part of spin structure function, $g_1^{NS}(x,Q^2)$ at small-$x$ by means of analytical solution of DGLAP equation in LO, NLO and NNLO using a Regge like ansatz with $Q^2$ dependent intercept as the initial input. The solutions are used to calculate the Bjorken integral $\Gamma_1^{p-n}(x_{min},Q^2)$ with QCD corrections up to NNLO. In addition, we have estimated the higher twist correction for the Bjorken sum rule.  We have performed phenomenological analyses of these results for $g_1^{NS}(x,Q^2)$ and $\Gamma_1^{p-n}(x_{min},Q^2)$ with different experimental data and model predictions. From the phenomenological analysis we have the following observations:

\ i. The Regge inspired ansatz in accord with DGLAP equations provides a very good description of the small-$x$ behaviour of $g_1^{NS}(x,Q^2)$, which are consistent with other results taken from Ref.~\cite{ppdf10,nmn41,nmn42,nmn43,nmn44,nmn51,nmn52,nmn53}.

\ ii. Our expressions for Bjorken Integral, along with considerable higher twist correction provide a very good description of the following experimental measurements: COMPASS \cite{nmn41}, HERMES\cite{nmn42}, E143\cite{nmn44} and JLab experiments \cite{BSRJLAB1,BSRJLAB2,BSRJLAB3}, which indicates that the experimental data strongly confirm the BSR prediction.

\ iii. Our results for $Q^2$ behaviour of Bjorken integral are also consistent with the QCD predictions up to NNLO. This consistency between our results and theoretical QCD predictions suggests that available data, the Regge ansatz and the theoretical framework of pQCD, through this simple method allow us to have a clean test of pQCD predictions on BSR.

\ iv. The consistency of the results for $xg_1^{NS}(x,Q^2)$ and $\Gamma_1^{p-n}$ due to the Regge like model, $xg_1^{NS}(x,t) = Ax^{1-b t}$  with different experimental results\cite{nmn41,nmn42,nmn43,nmn44,BSRJLAB1,BSRJLAB2,BSRJLAB3} and other strong analysis \cite{ppdf10,nmn51,nmn52,nmn53,BSR BI,BSR Soffer,KPSST,KS} signifies that the model is applicable in describing the small-$x$ behaviour of  $xg_1^{NS}(x,Q^2)$ structure function although it being simple. Moreover, in this method we do not require the knowledge of initial distributions of structure functions at all values of $x$ from 0 to 1. Here, we just require one input point at any fixed $x$ and $Q^2$ and with respect to that point both the $x$ and $Q^2$ evolution of structure functions can be obtained.

\ Our concluding impression based on all these observations is that the simple but efficient $Q^2$ dependent Regge ansatz for $xg_1^{NS}(x,Q^2)$ is capable of evolving successfully the  $xg_1^{NS}(x,Q^2)$ structure function in accord with DGLAP equation at small-$x$ and the Regge ansatz and the theoretical framework of pQCD, along with available experimental data lead towards a clean test of pQCD predictions of Bjorken Sum Rule.

%
%

%
%

\end{document}